\newif\ifPDF
\ifx\pdfoutput\undefined\PDFfalse
\else\ifnum\pdfoutput > 0\PDFtrue
     \else\PDFfalse
     \fi
\fi
\documentclass[usenatbib]{mn2e}
\ifPDF
  \usepackage[T1]{fontenc}
  \usepackage{aeguill}
  \usepackage[pdftex]{graphicx,color}
  \usepackage{subfigure}
  \usepackage{afterpage}
\else
  \usepackage[T1]{fontenc}
  \usepackage[dvips]{graphicx,color}
  \usepackage{subfigure}
  \usepackage{afterpage}
\fi

\title[Pulsar DM]{Tracking pulsar dispersion measures using the GMRT}
\author[A.L. Ahuja et al.] 
{A. ~L. ~Ahuja,$^1$ Y. Gupta,$^2$, D. Mitra,$^2$ and A. K. Kembhavi$^1$
\\
$^1$ IUCAA, Ganeshkhind, Pune University, Pune, India \\
$^2$ National Centre for Radio Astrophysics, TIFR, Pune University Campus, Pune 411007, India}
\date{Released 2004 Xxxxx XX}

\begin{document}
\maketitle


\begin{abstract}
In this paper, we describe a novel experiment for the accurate estimation of pulsar 
dispersion measures using the Giant Metre-wave Radio Telescope.  This experiment 
was carried out for a sample of twelve pulsars, over a period of more than one 
year (January 2001 to May 2002) with observations about once every fortnight.  
At each epoch, the pulsar DMs were obtained from simultaneous dual frequency 
observations, without requiring any absolute timing information.  The DM estimates 
were obtained from both the single pulse data streams and from the average profiles.  
The accuracy of the DM estimates at each epoch is $\sim$ 1 part in $10^4$ or better, 
making the data set useful for many different kinds of studies.  

The time series of DM shows significant variations on time scales of weeks to months
for most of the pulsars.  A comparison of the mean DM values from these data show 
significant deviations from catalog values (as well as from other estimates in 
literature) for some of the pulsars, with PSR B1642$-$03 showing the most notable 
changes.  From our analysis results it appears that constancy of pulsar DMs (at the 
level of 1 in $10^3$ or better) can not be taken for granted.  For PSR B2217$+$47, we
see evidence for a large-scale DM gradient over a one year period, which is modeled
as being due to a blob of enhanced electron density sampled by the line of sight.
For some pulsars, including pulsars with fairly simple profiles like PSR B1642$-$03, 
we find evidence for small changes in DM values for different frequency pairs of 
measurement, a result that needs to be investigated in detail.  Another interesting 
result is that we find significant differences in DM values obtained from average 
profiles and single pulse data. 
\end{abstract}


\section{Introduction }                 \label{sec:intro}
The radio signals from a pulsar suffer dispersion as they travel through 
the ionized component of the inter-stellar medium (ISM), resulting in a frequency 
dependent arrival time of the pulses.  The effect is quantified by the 
pulsar's dispersion measure (DM), defined as the integral of the electron 
column density along the line of sight, 
\begin{equation} 
DM ~~=~~ \int_{0}^{L} n_e dl ~~{\rm pc/cm^3}  ~~~.    \label{eqn:DM_defn}
\end{equation}
The delay between the pulse arrival time at two frequencies, $\Delta t$, can then 
be expressed as
\begin{equation} 
\Delta t ~~=~~  K ~{\left( \frac{1}{f_1^2} - \frac{1}{f_2^2}\right)} ~ DM ~~~, 
\label{eqn:dual_f_del} 
\end{equation}
where
\begin{equation} 
~~K = \frac{e^2} {2\pi m c} = \frac{1}{2.410331 \times 10^{-4}} \rm ~MHz^2~cm^3~s/pc ~~~.
\end{equation}
Here $\Delta t$ is in units of second for $f_1$ and $f_2$ in MHz and DM in the 
traditional units of $\rm pc/cm^3$.  The precise value of the constant K is as given
in \cite {Backer_etal}.

The DM of a pulsar is a basic parameter, and its value needs to be known with 
sufficient accuracy for proper dispersion correction to be carried out on 
the received signal.  Further, accurate estimates of DM can be used to probe 
the pulsar emission geometry \citep[e.g.][]{Kardashev_etal}.  Estimates of DM 
obtained from different values of $f_1$ and $f_2$ in Equation \ref{eqn:dual_f_del}
have been used to check the validity of the cold plasma dispersion relation 
for the ISM \citep[e.g.][and references therein]{PnW-92}.
In addition, small variations in a pulsar's DM are expected due to random electron 
density fluctuations in the ISM, thought to be associated with turbulence in the 
medium.  Such variations, expected on relatively large time-scales of weeks to months, 
have indeed been observed \citep[e.g.][]{Backer_etal,PnW-91}.  
Pulsar dispersion monitoring thus provides a direct method for probing the structure 
of the spectrum of electron density fluctuations. 

Though first order estimates of the DM can be obtained by careful measurements
of the arrival time delays in a multi-channel receiver operating at a single
wave-band (e.g. during the pulsar search and discovery process itself) 
the more accurate estimates needed for the applications discussed above require
more sophisticated experiments.  Typically, refined pulsar DMs (and their
variations with epoch) are estimated as part of the analysis of multi-epoch 
multi-frequency timing data from an observatory \citep[e.g.][]{Backer_etal, PnW-92}.  
An alternate method is to conduct simultaneous dual frequency 
observations at $f_1$ and $f_2$ and estimate the DM from a measure of the 
arrival time delay, using Equation \ref{eqn:dual_f_del} 
\citep[e.g.][]{Bartel_etal, Kardashev_etal, Hankins-87}.  The advantage of this 
method is that observations at a single epoch are self-sufficient for obtaining the 
DM at that epoch and the DM is obtained more directly, rather than as one of the 
parameters in a multi-parameter timing solution.  For single dish telescopes, this 
method requires simultaneous operation of receivers at more than one wave-band; 
alternatively, different single dish telescopes can be configured at each wave-band 
while simultaneously observing the same pulsar.

In this paper, we describe a new experiment for accurate estimation of pulsar DMs, 
using the Giant Metre-wave Radio Telescope (GMRT) in a simultaneous multi-frequency 
pulsar observation mode.  Section \ref{sec:strategy} describes the details of the
experiment and the observation strategy.  Section \ref{sec:dm_est} gives the
details of the data reduction, and describes the technique used for estimating 
DMs from the reduced data. The main results and the possibilities for follow-up
work are described in Section \ref{sec:results}.

\section{A new experiment for measuring DM}              
\label{sec:strategy}
\begin{table*}
\begin{minipage}{170mm}
\caption[Sample of pulsars]{Relevant parameters of our selected sample of pulsars.}
\label{tab:psrp}
\begin{tabular}{|c|r|r|r|r|r|c|c|} 
\hline
Pulsar & Catalog DM & Period & $\rm S_{400}$ & Distance & 
$V_{pm}$ & Duration of & Frequency combination \\ 
& ($\rm pc/cm^3$) & ($\rm sec$) & ($\rm mJy$) & ($\rm kpc$) & ($\rm km~s^{-1}$) &
scan ($\rm min$) & of observation ($\rm MHz$) \\ \hline
B0329$+$54 & 26.776  &  0.7145 & 1650 &  1.43 & 145 & 33 &227-243 $+$ 610-626 \\
B0818$-$13 & 40.99   &  1.2381 &  100 &  2.46 & 376 & 22 &227-243 $+$ 325-341 \\
B0823$+$26 & 19.4751 &  0.5307 &   65 &  0.38 & 196 & 22 &227-243 $+$ 325-341 \\
B0834$+$06 & 12.8579 &  1.2738 &   85 &  0.72 & 174 & 22 &227-243 $+$ 325-341 \\
B0950$+$08 &  2.9702 &  0.2531 &  400 &  0.12 &  21 & 22 &325-341 $+$ 610-626 \\
	 &	   &	     &	    &	    &	&     &  \\
B1133$+$16 &  4.8471 &  1.1877 &  300 &  0.27 & 475 & 33 &325-341 $+$ 610-626 \\
B1642$-$03 & 35.665  &  0.3877 &  300 &  2.90 & 660 & 11 &325-341 $+$ 610-626 \\
B1642$-$03 & 35.665  &  0.3877 &  300 &  2.90 & 660 & 11 &227-243 $+$ 325-341 \\
B1919$+$21 & 12.4309 &  1.3373 &  200 &  0.66 & 122 & 11 &227-243 $+$ 325-341 \\
B1929$+$10 &  3.176  &  0.2265 &  250 &  0.17 &  86 & 11 &227-243 $+$ 325-341 \\
	 &	   &	     &	    &	    &	&      &  \\
B1929$+$10 &  3.176  &  0.2265 &  250 &  0.17 &  86 & 22 &325-341 $+$ 610-626 \\
B2016$+$28 & 14.176  &  0.5579 &  320 &  1.10 &  12 & 11 &227-243 $+$ 314-320 \\
B2016$+$28 & 14.176  &  0.5579 &  320 &  1.10 &  12 & 22 &325-341 $+$ 610-626 \\
B2045$-$16 & 11.51   &  1.9616 &  125 &  0.64 & 289 & 11 &227-243 $+$ 314-320 \\
B2217$+$47 & 43.54   &  0.5385 &  135 &  2.45 & 375 & 22 &325-341 $+$ 610-626 \\
\hline
\end{tabular}
\medskip
\end{minipage}
\end{table*}
The accuracy of the DM estimate depends on the precision to which the
the time delay between the pulse profiles at two frequencies can be
measured.  If $\Delta t_{rms}$ is the error on the measurement of the time delay,
then the fractional DM error is
\begin{equation} 
\frac{DM_{rms}}{DM} ~~=~~ \frac{\Delta t_{rms}}{\Delta t}  ~~~.    \label{eqn:dm_acc} 
\end{equation}
For a given value of $\Delta t_{rms}$ (which is usually limited by the S/N of the 
data at the two frequencies, or sometimes by the coarseness of the sampling interval), 
it is clear that the greater the relative time delay between the arrival of signals 
at the two frequencies, $\Delta t$, the more accurate is the DM estimate.  
This would favour large separations between the two observing radio bands.  
However, if the pulsar profile evolves significantly over this range of
frequencies, then it can bias the measured $\Delta t$, leading to an error in the
estimate of the DM. This effect favours a smaller separation between the two radio 
wave-bands. Also, according to Equation \ref{eqn:dual_f_del}, for a given 
separation between a pair of radio bands $f_1$ and $ f_2$, smaller values of frequencies 
give a larger value of estimated $\Delta t$, and in turn, a better accuracy for 
final DM estimation. The final choice of the two frequency bands of operations is 
then decided by these considerations. Other requirements for
obtaining accurate DM estimates are (a) high signal to noise ratio stable pulse 
profiles, which are more readily observed at low radio frequencies (typically in 
the range 100 to 1000 MHz) where the pulsar is known to be bright and 
(b) accurate time alignment of the multi-frequency pulse profiles.
As we now describe, the GMRT, because of some unique features, offers a novel way 
for obtaining accurate DM estimates.  

The GMRT is a multi-element aperture synthesis telescope \citep[][]{Swarup_etal} 
consisting of 30 antennas, distributed over a region of 25 km diameter, 
which can also be configured as a ``single dish'' in the incoherent or coherent 
array mode \citep[][]{gupta_etal_00}.  Furthermore, it supports a ``sub-array'' 
mode of operation where different sets of antennas can be configured completely 
independently to produce more than one single dish.  Thus, the same pulsar can 
be observed simultaneously at more than one radio band.  

The GMRT operates at radio frequencies in the range 150 MHz to 1400 MHz with 
observing bands available at 150, 235, 325, 610 and 1400 MHz. 
The antennas can be grouped into several sub-arrays and each sub-array 
can independently be operated at a radio band of interest, thus enabling
simultaneous multi-frequency observations. Signals from different observing
frequency bands and antennas are eventually down-converted to baseband signals
of 16 MHz band-width. The signals are subsequently sampled at the Nyquist rate 
and processed through a digital receiver system consisting of a correlator and
a pulsar back-end.

For each antenna, operating at a given frequency band, the pulsar back-end receives 
signals in 256 channels spanning the band-width of 16 MHz, for each of two orthogonal 
polarizations.  The relative delay -- geometrical as well as instrumental -- between 
different antenna signals is compensated to an accuracy of 32 nanosec before they 
reach the pulsar receiver.  The corresponding signals from selected antennas (say 
from one sub-array) can be added together incoherently by the pulsar receiver.  

For this experiment, the signals from antennas in all sub-arrays were added incoherently 
in the same pulsar receiver, to produce a single stream of output data, which was 
recorded at a sampling rate of 0.516 millisecond.  Because of the dispersive delay 
between the different radio bands of observation, the pulse arrives at different 
times (and hence, at different pulse phases) at each frequency band.  This fact is 
utilised to separately extract the streams of single pulses at each frequency band, 
from the single stream of recorded data, during the offline analysis.  This scheme 
eliminates the need for having separate, but synchronised, pulsar receiver chains 
for each sub-array and also does away with any requirement of accuracy of absolute 
time stamping of the recorded data -- the data from the different sub-arrays is 
naturally synchronised.  Since all known instrumental and geometric delays have 
been corrected for all the sub-arrays, the residual arrival time delay between 
pulses from different radio bands of observation is only and entirely due to the 
dispersion delay.  This allows the DM to be measured to a very high degree of accuracy.
%
\begin{figure}
\begin{center}
\includegraphics[angle=-90, width=0.45\textwidth]{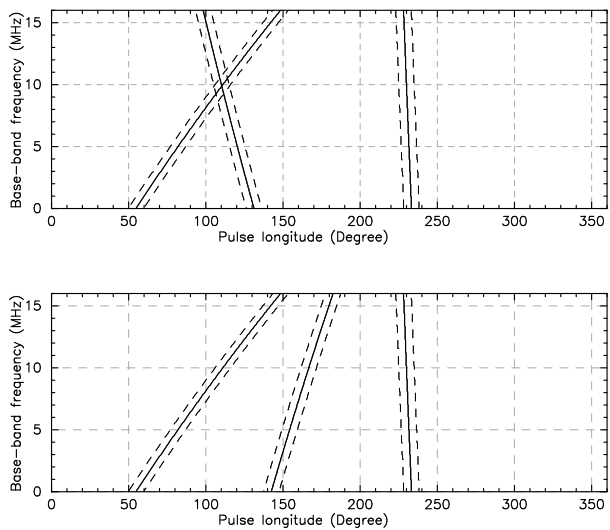}
\caption[DM delay tracks]{Dispersion curves across the 16 MHz of base-band signal 
for pulsar B2016$+$28. The upper panel shows (left to right) the dispersion curves 
for the 243 to 227, 325 to 341 and 610 to 626 MHz bands of observation \& the lower 
panel shows the dispersion curves for 243 to 227, 320 to 304 and 610 to 626 MHz 
bands, respectively. The dotted curves on both sides of the continuous curves 
delineate the extent of the 50\% width of the average profile.}
\label{fig:dm_delay}
\end{center}
\end{figure}

There is, however, one drawback of the above scheme.  In order to recover the 
pulsed signal for the different frequency bands during off-line analysis, dispersion 
delays across the 256 channels (16 MHz baseband band-width) for each frequency 
band are computed and the data are collapsed to obtain a time series for each band.
In this process, however, the data from the other frequency band are wrongly 
de-dispersed and appear as a smeared out signal producing excess undesired power 
in the off-pulse region. In some cases this may overlap with the on-pulse signal
from the desired frequency band, resulting in corruption of the data.  Thus, in order 
to obtain undistorted signals, it is essential that we choose an observing strategy 
that avoids such overlaps.  This requires us to examine the detailed nature of the 
DM delay curve at each frequency band of interest, and to ensure that the curves 
do not intersect each other within the 16 MHz of baseband band-width. In Figure 
\ref{fig:dm_delay} we show an example of this. Here, the upper panel (from left to 
right) shows the dispersion curves for pulsar B2016$+$28 in the frequency bands 
243 to 227, 325 to 341 and 610 to 626 MHz, as seen in the base-band signal, after 
removal of all delays that are integer multiple of the pulsar period. It shows that 
the two dispersion curves at frequency bands 243-227 MHz and 325-341 MHz intersect with one 
another over certain range of channels.  Hence, this combination of frequency bands 
can not be used for such observations of this pulsar.  By suitably changing the value 
of the local oscillator signals used for the down conversion of the radio frequency 
bands to base-band signals, the range as well as the direction of the radio frequency 
signals that span the 16 MHz band-width can be changed, thus ensuring proper separation 
of the dispersion curves.  In this particular case, it has been achieved by moving the 
local oscillator such that the 325 MHz band covers 320 to 304 MHz (see lower panel 
of Figure \ref{fig:dm_delay}).  Appropriate frequency combinations were found for 
each pulsar in our sample. 

For this experiment, we selected a sample of 12 pulsars having sufficiently large fluxes 
($S_{400} > 100 ~\rm mJy $), a range of DM values ($\sim 10-40 ~\rm pc/cm^3$), 
and sampling different directions in the Galaxy.  The relevant parameters 
are summarised in Table \ref{tab:psrp}, where columns 2,3,4,5 and 6 give the 
values of the DM, period, flux at 400 MHz, distance and proper motion
respectively, as obtained from the pulsar catalog of \cite{cat-93}.  At every 
epoch of observation, each pulsar from our sample was observed for a few thousand 
pulses (column 7 gives the duration of the observing scan) at a pair of frequency 
bands (given in column 8 of Table \ref{tab:psrp}) selected from the available
 bands of the GMRT.  The epochs were separated by intervals of about two 
weeks, and the whole experiment was carried out over a duration of about one and 
half years.
\section{Data reduction and estimation of DM} 	
\label{sec:dm_est}
The recorded data were pre-processed off-line to convert from raw 
time-frequency format to a single pulse time series and folded 
profiles. The pre-processing involved de-dispersion 
of the data in two frequency bands, folding and interference rejection.  

For each pulsar, to recover the pulse trains at the two radio bands, the 
acquired data were de-dispersed within the 16 MHz band-width of each band 
by using the catalog DM values given in Table \ref{tab:psrp}.
Where needed, bad data points were rejected from the de-dispersed data.  For this,
after masking the data from the on-pulse regions, the running mean data from the
off-pulse regions was computed and subtracted from the original data.  Next, off-pulse
data points with amplitude
greater than the threshold value (typically chosen as 3 times the off-pulse RMS) were 
flagged. In addition, data were scanned visually, and manual editing of bad data due 
to radio frequency interference was
carried out, where needed. At the end of the data rejection step, if a large 
fraction of the data around any on-pulse window was found to be bad, the entire 
pulse was flagged.

\begin{figure}
\begin{center}
\includegraphics[angle=-90, width=0.5\textwidth]{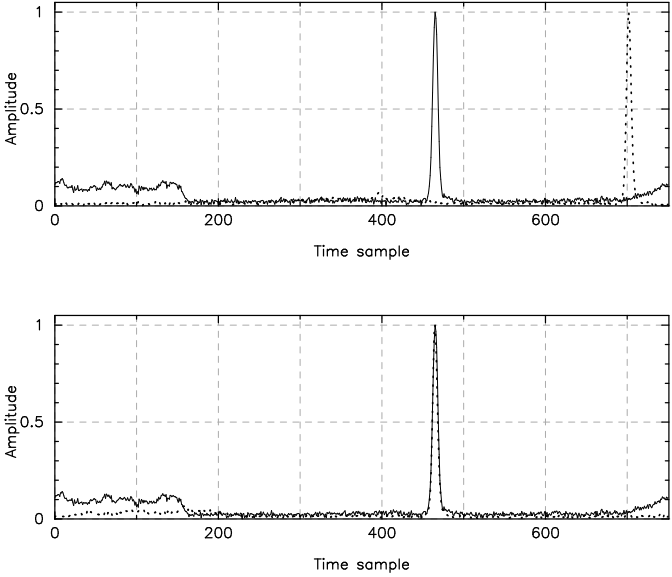}
\caption[Average pulse profile of B1642$-$03]{Average pulse profiles of the 
pulsar B1642$-$03 observed at 610 (solid curve) $+$ 325 (dotted curve) MHz bands
combination.  The upper and lower panels show the pulse profiles before and 
after the alignment respectively. The excess power regions near both edges 
of the profile at 610 MHz are examples of wrongly de-dispersed data from the other
band.}
\label{fig:prof}
\end{center}
\end{figure}

The de-dispersed, interference free data trains were folded at the Doppler-corrected 
pulsar periods to obtain the average pulse profiles at the two radio 
frequency bands (see Figure \ref{fig:prof} for an example).  The pulse profile 
data at each observation band were demarcated with three windows $-$ two 
off-pulse and one on-pulse window. The on-pulse window contained the properly 
de-dispersed average pulse profile, while the off-pulse windows (one on each side 
of the on-pulse) were off-pulse regions which were free of contamination from 
the wrongly de-dispersed pulse profile of the other frequency band.  Data only 
from these window regions were used in the subsequent analysis described below.

%
From the reduced data, the dispersion delay between the two frequency bands 
was estimated and, using Equation \ref{eqn:dual_f_del}, the corresponding DM
value was obtained.  For these calculations, Doppler corrected frequencies 
$f_1$ and $f_2$ (with $f_{1} > f_{2}$) were used, with these frequencies being 
related to the frequencies of observations, $f_{1m}$ and $f_{2m}$, through
\begin{equation} 
f_{1} = f_{1m} \sqrt{\frac{1 + \beta}{1 - \beta}} 
~~and~~ f_{2} = f_{2m} \sqrt{\frac{1 + \beta}{1 - \beta}}, 
~~ \beta = \frac{v_{net}}{c} ~~; \label{eqn:f_corr} 
\end{equation} 
where $v_{net}$ is the radial velocity of the observer with respect to the 
pulsar, which is predominantly due to the orbital motion of the earth around 
the Sun.  Similarly, the value of $\Delta t$ in Equation \ref{eqn:dual_f_del} 
needs to be the measured topocentric delay, $\Delta t_{m}$, corrected to the 
solar system barycenter, as follows: 
\begin{equation} 
\Delta t = \Delta t_{m} \times \left( 1 - \beta \right) ~~~.   \label{eqn:t_corr} 
\end{equation}
The total measured time delay, $\Delta t_{m}$, can be expressed as a sum of three terms: 
\begin{equation} 
\Delta t_{m} = \Delta t_p + \Delta t_i + \Delta t_f ~~~,
\end{equation}
where $\Delta t_p$ is the integral number of pulsar periods delay, $\Delta t_i$ 
is the number of time sample bins delay within a pulsar period and $\Delta t_f$ 
is the fraction of a time sample bin delay.
The value of $\Delta t_{m}$ can be estimated by two different techniques: (i) by 
estimating the delay between the average pulse profiles, and (ii) by measuring 
the mean delay between the single pulse data trains.  We have carried out the 
analysis using both these methods, and the steps for each are described below.

As the first step, the data were reduced to zero mean off-pulse sequences.
In the average profile (hereafter AP) method, the mean from the off-pulse data 
windows was estimated and subtracted from the whole pulse profile data.  In the 
single pulse (hereafter SP) method, the mean computation and baseline subtraction 
was carried out individually for each pulse, while using the same off-pulse windows.

In the AP method, because of the folding process, the value of $\Delta t_p$ can not
be directly estimated from the folded profiles; instead, it was estimated from the
knowledge of the frequencies for the two bands, the catalog DM value and the pulsar
period.  To estimate $\Delta t_i$, pulse profiles at the two frequency bands were 
cross-correlated, and the integer time sample lag at which the cross-correlation 
peaked was taken as value of $\Delta t_i$. The lower frequency pulse profile was 
rotated left circularly by this amount to align it with the higher frequency pulse 
profile (see Figure \ref{fig:prof} for an example).  

\begin{figure}
\begin{center}
\includegraphics[angle=-90, width=0.5\textwidth]{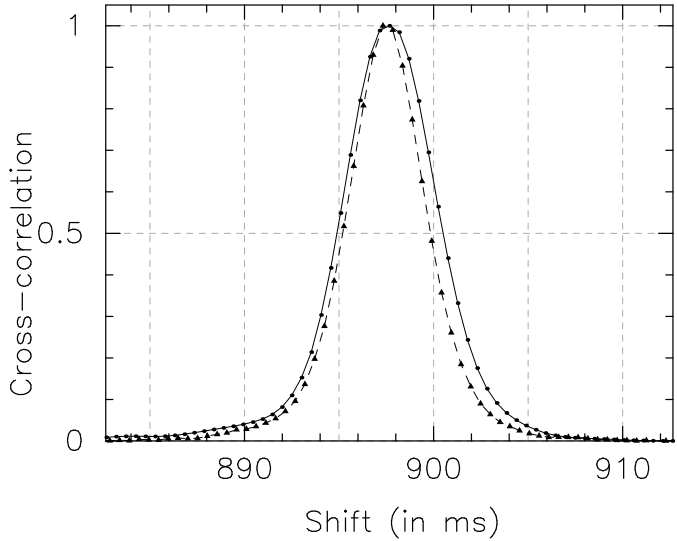}
\caption[Cross-correlation function for pulsar B1642$-$03]{The normalized cross-correlation
function (CCF) for pulsar B1642$-$03 observed at 610$+$325 MHz bands. 
The continuous curve shows the CCF for average pulse profiles and dashed one 
correspond to single pulse analysis.} 
\label{fig:ccf}
\end{center}
\end{figure}

The cross-correlation (hereafter 
CC) of the pulse profiles at two given frequencies (see Figure \ref{fig:ccf} for an
example) can be given as, 
\begin{equation} 
CC(kT) ~~=~~ \sum_{n=1}^N f(nT) g(nT-kT). 
\end{equation}
Here, CC(kT) is the CC for $k^{th}$ bin shift of the pulse profile at the lower 
frequency, N is the number of time sample bins within an on-pulse window, 
and $f$ and $g$ are the pulse profiles at two observation frequencies.
%
%
In the SP method, the two time series were cross-correlated, and the 
peak of cross-correlation function gave the time delay with an accuracy of 
a time sample bin. In this method, the CC could be started from zero 
shift of the lower frequency pulse profile, but to reduce unnecessary 
computations, we started CC computations from a shift equivalent to the 
number of time sample bins corresponding to $\Delta t_p$.
During the cross-correlation computations in both the methods, care was 
taken to ensure that data points from the wrongly de-dispersed signals were 
not included in the computations. This was done by using data points from
the above defined on-pulse and off-pulse windows only, and restricting the 
lag range to values which ensured no overlap of these windows with wrongly 
de-dispersed data points. 

The average profile is obtained by folding the time series data at the pulsar 
period.  Since individual pulses show significant pulse to pulse jitter in the 
longitude of occurrence, the average profile is usually significantly broader 
than the individual pulses. As a result the CCF obtained in the AP analysis is 
broader in comparison to that from the SP analysis (e.g. Figure \ref{fig:ccf}).
In the AP method, the CCF reflects the sum of cross-correlation of all pulses at 
one radio band with all pulses from the other band, while in the SP analysis, 
the CCF is the sum of the CC between corresponding pulses at the two radio bands. 
Therefore, one can expect the DM delay estimated by the two methods to be 
different, as we find in our results. 

The precision of DM measurement mainly depends on the accuracy in estimating
the time delay between two pulse profiles. The CC as described above gives an 
accuracy of the order of an integral time sample bin. To estimate the delay 
with an accuracy of a fraction of a time sample bin, the cross-spectrum (CS) 
was computed and a linear gradient was fitted to the phase of the CS. Let us 
first consider the AP method. If the two pulse profiles are $f(t)$ and $g(t)$, 
then their Fourier transforms (FT) can be written as,
\begin{equation} 
f(t) \Longleftrightarrow F(\nu) ~~=~~ \arrowvert 
F(\nu)\arrowvert e^{i \left(\phi_{1i} + 2\pi \nu t_{1f}\right)}
\end{equation}
and
\begin{equation} 
g(t) \Longleftrightarrow G(\nu) ~~=~~ \arrowvert 
G(\nu)\arrowvert e^{i \left(\phi_{2i} + 2\pi \nu t_{2f}\right)} ~~~; 
\end{equation}
where $\arrowvert F(\nu)\arrowvert$ and $\arrowvert G(\nu)\arrowvert$ are 
the amplitudes of Fourier transform components at the transform frequency 
$\nu$, $t_{1f}$ and $t_{2f}$ are the positions of the peaks of the two 
pulse profiles from their reference points of Fourier transformation in the 
time domain, and $\phi_{1i}$ and $\phi_{2i}$ are the intrinsic phases of the 
two pulse profiles.  The cross-spectrum can then be written as,
\begin{equation} 
CS(\nu) = F(\nu)G^*(\nu) ~~=~~ |F(\nu)||G(\nu)|e^{-i\phi_{CS}\left(\nu\right)} ~~~;
\end{equation}
where the phase $\phi_{CS}\left(\nu\right)$ is given by
\begin{equation} 
\phi_{CS}\left(\nu\right) ~~=~~ \phi_{2i} - \phi_{1i} + 2\pi\nu\Delta t_f ~~~, 
\end{equation}
with $\Delta t_f ~=~ t_{2f} ~-~ t_{1f}$ the fractional time 
sample bin delay.
  
\begin{figure}
\begin{center}
\includegraphics[angle=-90, width=0.5\textwidth]{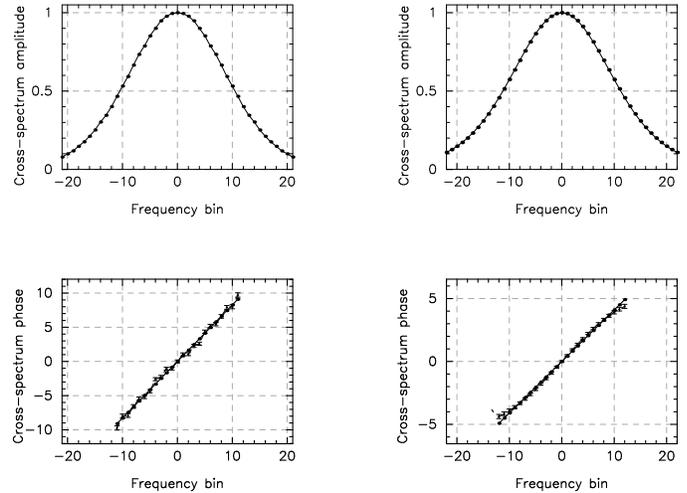}
\caption[AP CS BSTFT line]{Normalized CS amplitude (upper panels), and 
CS phase with error bars (lower panels) of average profiles (left side panels) 
and single pulses (right side panels) for pulsar B1642$-$03, at one epoch 
observed at 610$+$325 MHz bands. The straight line in the phase plot is the 
best fit linear gradient.}
\label{fig:bstft}
\end{center}
\end{figure}

For $\phi_{1i}(\nu) = \phi_{2i}(\nu)$, i.e. when the pulse profiles at the two 
frequencies have the same shape, the effect of a fractional bin delay will show 
up as a linear gradient in the phase plot of the CS (see Figure \ref{fig:bstft}), 
given by
\begin{equation} 
\Delta t_f = \frac{\Delta\phi_{CS}}{2\pi\Delta\nu} ~~~. 
\end{equation}
The cross-spectrum can be obtained from the Fourier transformation of the CCF 
or from the product of the individual Fourier transformations.  Of the two we 
have preferred the latter for the AP method, as this helps in the proper 
propagation of errors from time domain to frequency domain, as explained 
below.  In the SP analysis, however, we have used the Fourier transformation 
of the CCF,
with an appropriate strategy for computing the errors in the final DM results.

Let us now look at the estimation of the error in the measured delay, 
which is primarily due to the finite signal to noise ratio of the data. 
For the AP method, the noise in the folded profiles, estimated from the 
off-pulse windows, was properly propagated to the CS. For each pulse profile, 
the RMS of phase, $\sigma_{\phi\left(\nu\right)}$, and amplitude,
$\sigma_{A\left(\nu\right)}$, of the Fourier transform can be estimated as, 
\begin{equation} 
\sigma_{\phi\left(\nu\right)} ~~=~~ \sigma_t \sqrt{\frac{N}{2 
\left(Im_{\nu}^2 + Re_{\nu}^2\right)}}, \label{eqn:fft_ph_rms} ~~~,
\end{equation}
\begin{equation} 
\sigma_{A\left(\nu\right)} ~~=~~ \sigma_t \sqrt{\frac{N}{2}} ~~~.
\end{equation}
Here $N$ is the number of data points used for fast Fourier transformation (FFT), 
$Im_{\nu}$ and $Re_{\nu}$ are real and imaginary parts respectively of the Fourier 
components at frequency bin $\nu$, and $\sigma_t$ is the RMS of the off-pulse noise.  
The RMS of the CS phase, $\sigma_{\phi_{CS}\left(\nu\right)}$, was computed by 
adding the noise from the two phases in quadrature, 
\begin{equation} 
\sigma_{\phi_{CS}\left(\nu\right)}^2 ~~=~~ \sigma_{\phi_1\left(\nu\right)}^2 + 
\sigma_{\phi_2\left(\nu\right)}^2 ~~~.
\end{equation}

In the SP method, the RMS obtained from off-pulse windows was properly propagated 
to estimate the RMS at each point of the CCF. The greatest value of this RMS was
used as a conservative estimate in Equation \ref{eqn:fft_ph_rms} to estimate the 
RMS of the CS phase.  
After this step, the procedure for estimating the error in the DM was the same 
for the AP and SP methods.

\begin{figure}
\begin{center}
\includegraphics[angle=-90, width=0.45\textwidth]{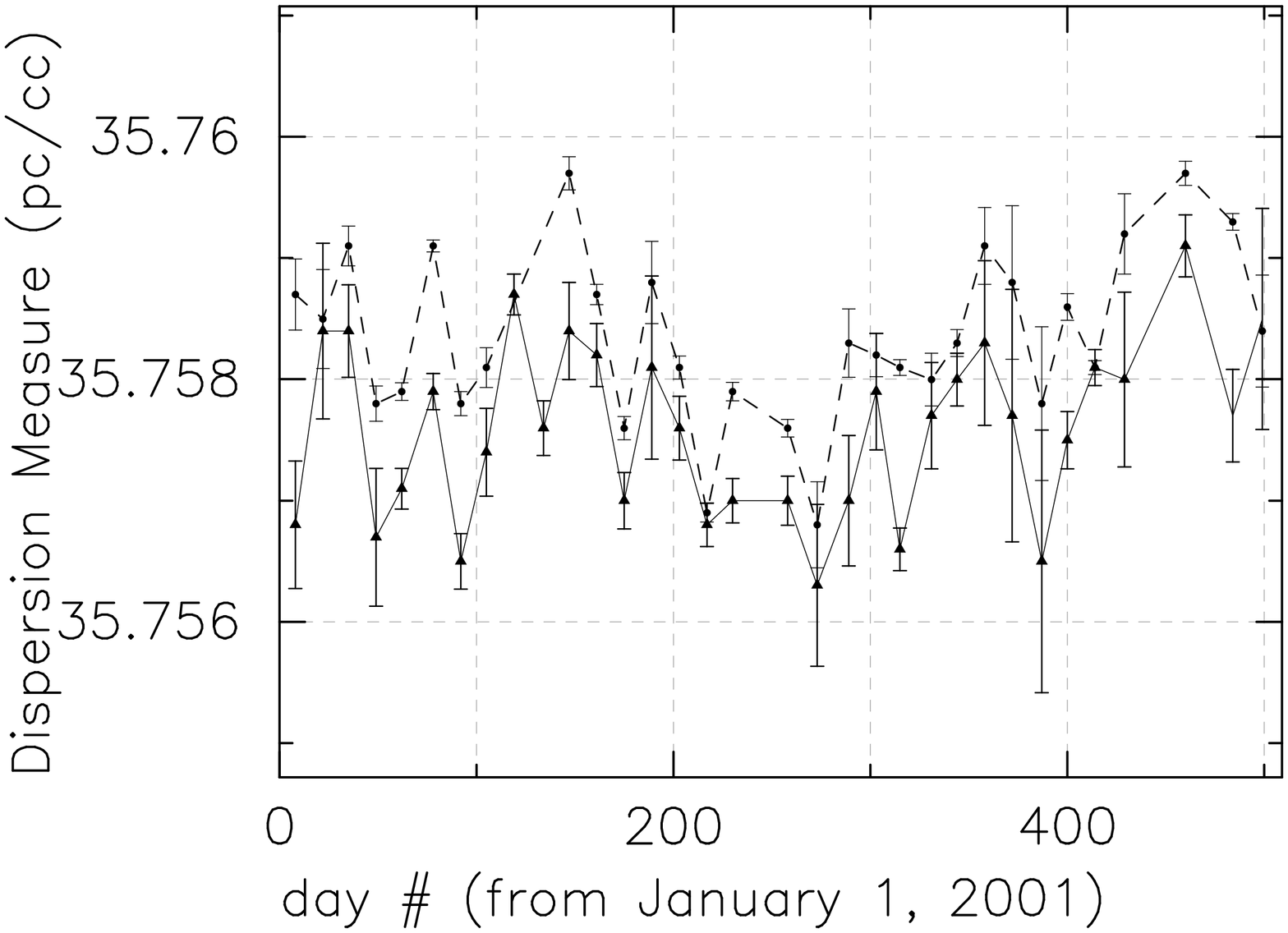}
\includegraphics[angle=-90, width=0.45\textwidth]{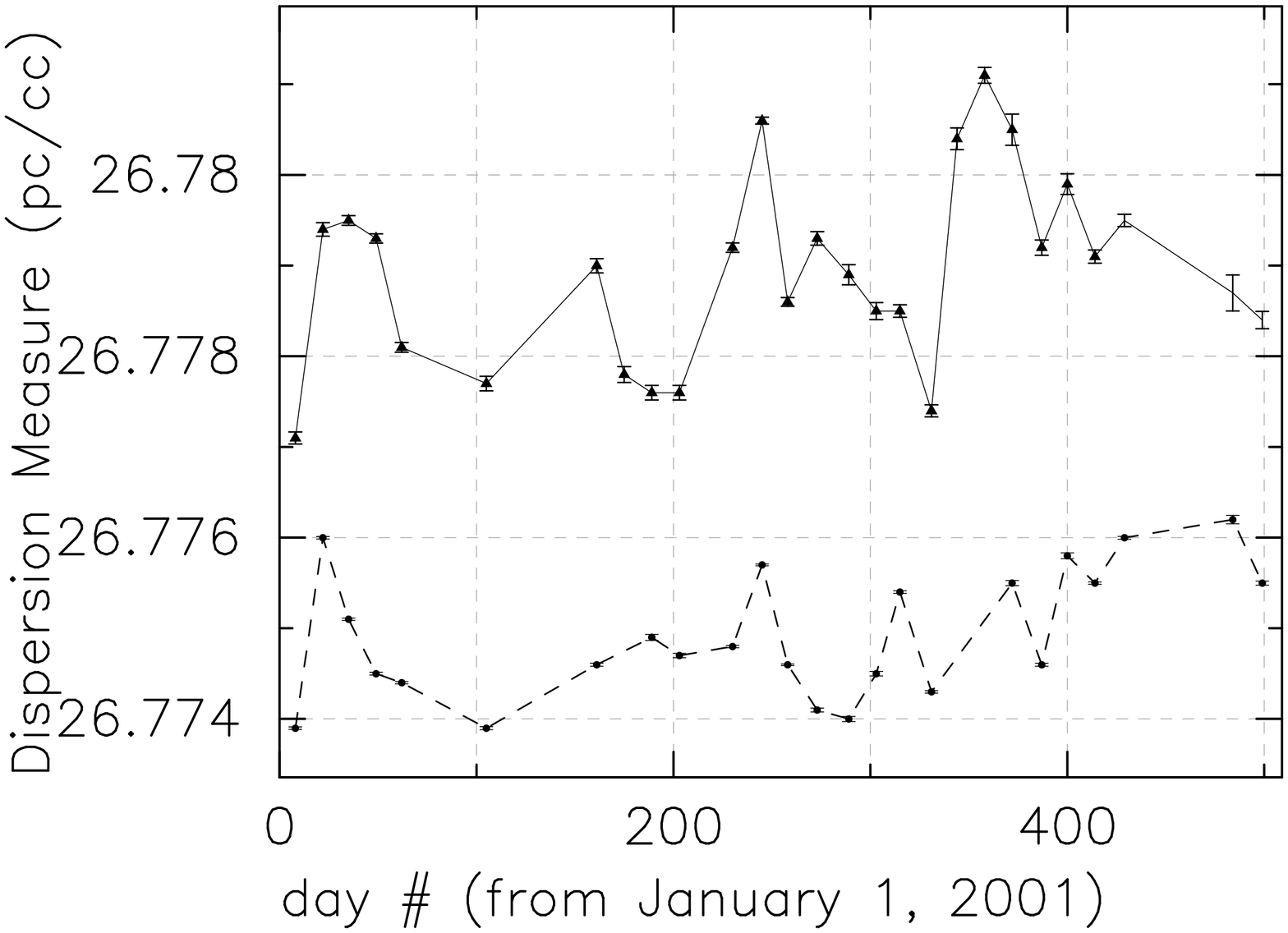}
\caption[Final results for pulsar B1642$-$03]{Variation of DM with time for pulsars 
B1642$-$03 (upper panel) and B0329$+$54 (lower panel) observed at frequencies 
610$+$325 MHz and 610+243 MHz respectively, over the interval 08 Jan 2001 to 
14 May 2002, as a function of day number. 
The continuous line shows the results from average profile analysis, and the 
dotted one from single pulse analysis.  The error bars are 3$\sigma_{DM_{(noise)}}$ 
values. }
\label{fig:dm_tm_srs}
\end{center}
\end{figure}

The phase gradient, $\nabla\left(\phi_{CS}\right)$, was computed as the slope 
of the best fitted line, $\nabla\left(\rm best fit\right)$, 
obtained by the least-square method.  Thus,
\begin{equation} 
\Delta t_f ~~=~~ \frac{\nabla\left(best fit\right) \times N_{FFT} \times T}{360} ~~~, 
\end{equation}
where $N_{FFT}$ is the number of data points used to compute the FFT and $T$ is 
the time sample. The RMS of $\Delta t_f$ was estimated as
\begin{equation} 
\sigma_{\Delta t_f} ~~=~~ \frac{\sigma_{\nabla\left(\rm best fit\right)}
\times N_{FFT} \times T}{360} ~~~.
\end{equation}
Because the error in $\Delta t$ estimation was only due to $\sigma_{\Delta t_f}$,
the error in the final DM value was given by
\begin{equation} 
\sigma_{DM_{\left(noise\right)}} ~~=~~ \frac{\sigma_{\Delta t_f}} {\Delta t_c} DM ~~~.
\end{equation}
The above steps were carried out at each epoch to obtain a time series
of DM values for each pulsar (see Figure \ref{fig:dm_tm_srs} for example).

\section{Results and discussions} 
\label{sec:results}

\begin{table*}
\begin{minipage}{190mm}
\caption{DM results from average profile analysis}
\label{tab:dm_res}
\begin{tabular}{|c|r|c|c|r|r|c|l|} \hline
Pulsar & Catalog DM & Frequency & $N_{ep}$ & $\langle DM\rangle$ & 
$\sigma_{DM_{\left(noise\right)}}$ & $\sigma_{DM_{\left(total\right)}}$
& $\Delta DM / \sigma_{DM_{\left(total\right)}}$ \\ 
 
& ($\rm pc/cm^3$) & combination (MHz) &     & ($\rm pc/cm^3$) & 
($\rm pc/cm^3$) & ($\rm pc/cm^3$) & ($\rm pc/cm^3$) \\  \hline
B0329$+$54  & 26.776  & 243 + 610& 26 &   26.77870 & 0.00003  & 0.00103  & $+$  2.64\\
B0818$-$13  & 40.99   & 243 + 325& 32 &   40.9222 & 0.0013  & 0.0043  & $-$ 15.71\\
B0823$+$26  & 19.4751 & 243 + 325& 29 &   19.4545 & 0.0004  & 0.0016  & $-$ 12.85\\
B0834$+$06  & 12.8579 & 243 + 325& 29 &   12.8671 & 0.0004  & 0.0017  & $+$  5.38\\
B0950$+$08  &  2.9702 & 325 + 610& 31 &    2.9597 & 0.0008  & 0.0050  & $-$   2.1\\
            &         &		      &		  & &          &          \\
B1133$+$16  &  4.8471 & 325 + 610& 34 &    4.8288 & 0.0006  & 0.0071  & $-$  2.57\\
B1642$-$03  & 35.665  & 325 + 610& 33 &   35.75760 & 0.00014  & 0.00072  & $+$128.20\\
B1642$-$03  & 35.665  & 243 + 325& 34 &   35.72270 & 0.00007  & 0.00090  & $+$ 64.00\\
B1919$+$21  & 12.4309 & 243 + 325& 32 &   12.4445 & 0.0011  & 0.0054  & $+$  2.50\\
B1929$+$10  &  3.176  & 243 + 325& 31 &    3.1755 & 0.0004  & 0.0015  & $-$  0.31\\
            &         &		      &		&	     	 &          &	   \\
B1929$+$10  &  3.176  & 325 + 610& 27 &    3.1750 & 0.0004  & 0.0020  & $-$  0.51\\
B2016$+$28  & 14.176  & 243 + 320& 29 &   14.1611 & 0.0007  & 0.0025  & $-$  6.07\\
B2016$+$28  & 14.176  & 325 + 610& 30 &   14.1664 & 0.0008  & 0.0051  & $-$  1.90\\
B2045$-$16  & 11.51   & 243 + 320& 31 &   11.5094 & 0.0012  & 0.0114  & $-$  0.05\\
B2217$+$47  & 43.54   & 325 + 610& 31 &   43.5196 & 0.0007  & 0.0061  & $-$  3.38\\ \hline
\end{tabular}
\end{minipage}
\end{table*}
\begin{figure}
\begin{center}
\includegraphics[angle=-90, width=0.45\textwidth]{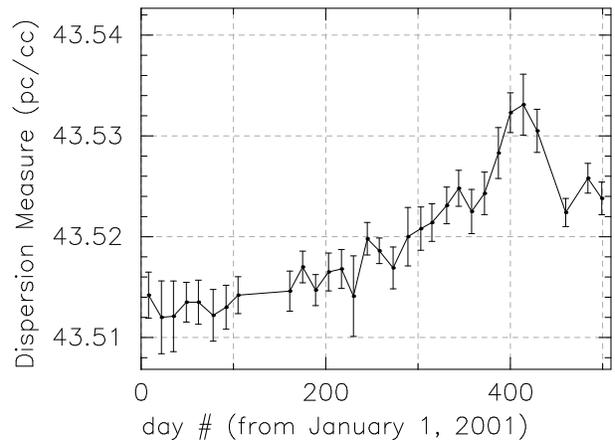}
\caption[Final results for pulsar B2217$+$47]{DM variation with 
3$\sigma_{DM_{(noise)}}$ 
error bars for pulsar B2217$+$47 observed at frequencies 610$+$325 MHz, over the 
time interval 08 Jan 2001 to 14 May 2002 as a function of day number.  
The catalog value of the DM is 43.54 $\rm pc/cm^3$.}
\label{fig:p2217}
\end{center}
\end{figure}
%
The results obtained for the average profile method are summarised in Table 
\ref{tab:dm_res}.  Here, column 2 gives the catalog DM value for each pulsar 
from \cite{cat-93}, and the observing frequency bands are given in column 3. 
For each pulsar, we obtained the mean dispersion measure over the period of 
observations, $\langle DM\rangle$, and the quadrature average of 
$\sigma_{DM_{\left(noise\right)}}$, using:
\begin{equation}  
\langle DM\rangle ~~=~~ \frac{\sum_{i=1}^{N_{ep}} DM_i}{N_{ep}} ~~~,  \label{eqn:avg_dm}
\end{equation}
\begin{equation} 
\sigma^2_{DM_{\left(noise\right)}} ~~=~~ 
\frac{\sum_{i=1}^{N_{ep}} \sigma^2_{DM_{i\left(noise\right)}}}{N_{ep}} ~~~; 
\label{eqn:sig_noise_dm}
\end{equation}
where $DM_i$ and $\sigma_{DM_{i\left(noise\right)}}$ are the measured dispersion 
measure and the RMS dispersion at the $i^{th}$ epoch, and $N_{ep}$ is the 
total number of epochs of observations (column 4 of Table \ref{tab:dm_res}). The 
quantity $\sigma_{DM_{\left(noise\right)}}$ (column 6 of Table \ref{tab:dm_res}) 
gives the average of the DM error bar estimate from all epochs of observations. 
This quantity gives an estimate of the contribution to the total RMS fluctuation 
seen in the time series, due to sources of error in the DM estimate.
The values for $\sigma_{DM_{\left(noise\right)}}$ for most pulsars are such that 
the DM estimate is accurate to 1 part in $10^4$ or better.  

We also estimated the total fluctuation of the DM time series, 
$\sigma_{DM_{\left(total\right)}}$ (column 7 of Table \ref{tab:dm_res}), as
\begin{equation} 
\sigma_{DM_{\left(total\right)}} ~~=~~ 
\sqrt{\frac{\left(\sum_{i=1}^{N_{ep}}  \left(DM_i ~-~ \langle DM\rangle\right)^2\right)}{N_{ep}}} ~~~. 
\label{eqn:sig_total_dm}
\end{equation}
In the most general case, this total RMS of the DM fluctuation is composed of a 
part due to estimation error on the DM (Equation \ref{eqn:sig_noise_dm}) and the 
remaining due to other processes likely to play a role in the time variability 
of DM (a prime candidate for which is DM fluctuation due to large scale electron 
density irregularities in the ISM).  An estimate of the variance due to such processes 
can be obtained as
\begin{equation} 
\sigma^2_{DM_{\left(ISM\right)}} ~~=~~ 
\sigma^2_{DM_{\left(total\right)}} ~-~ \sigma^2_{DM_{\left(noise\right)}} ~~~. 
\end{equation}
\label{eqn:sig_ISM_dm}
As can be seen in columns 6 and 7 of Table \ref{tab:dm_res}, for almost all the 
pulsars, $\sigma_{DM_{\left(noise\right)}}$ is much smaller than 
$\sigma_{DM_{\left(total\right)}}$, indicating the presence of substantial DM 
fluctuations due to such sources.  We return to this aspect in more detail at the
end of this section.

\subsection{On the constancy of $\langle DM \rangle$ estimates} 
\label{subsec:mean_DMs}

Keeping in mind the total RMS for each DM estimate ($\sigma_{DM_{\left(total\right)}}$),
we can see that the mean DM, $\langle DM\rangle$, for each pulsar is estimated with a fairly good
accuracy -- $\sim 1$ part in $10^3$ or better 
(DM accuracy at each epoch is $\sim 1$ part in $10^4$).  
It is interesting to compare these
mean DM values with other estimates in literature.  Column 8 of Table \ref{tab:dm_res} 
shows the difference between our $\langle DM\rangle$ value and the catalog DM value \citep[][]{cat-93}, 
in units of $\sigma_{DM_{\left(total\right)}}$.  While for most pulsars our results 
agree with the catalog values within 3 $\sigma_{DM_{\left(total\right)}}$, 
there are some pulsars, namely B0818$-$13, B0823$+$26, B0834$+$06, B1642$-$03 and 
B2016$+$28, which show a significant difference.  We now discuss these discrepant cases 
in some detail, using for comparison results from (i) the old pulsar catalog of 
\cite{cat-93} (ii) the new pulsar catalog \citep[][see also 
www.atnf.csiro.au/research/pulsar/psrcat]{Hobbs_etal}, and (iii) other reports in literature.

For pulsar B0818$-$13 we find a $\langle DM \rangle$ of $\rm 40.922 \pm 0.004 ~pc/cm^3$, which 
is significantly smaller than the value of $\rm 40.99 \pm 0.03 ~pc/cm^3$ given in the old
catalog, which comes from a very early measurement \citep[][]{MnT-72}.  It is 
interesting to note that \cite{kuz_etal_98} find an intermediate value of 
$\rm 40.965 ~ pc/cm^3$ for this pulsar, from measurements made between 1984 and 1991.  
Furthermore, the new pulsar catalog gives a value of $\rm 40.938 \pm 0.003 ~pc/cm^3$ 
\citep[][]{Hobbs_etal}, which is 
intermediate between that of \cite{kuz_etal_98} and our result.  One interesting
possibility from the above data points is that the DM of this pulsar is showing 
a slow and secular decline with time, on time scales of decades.

For pulsar B0823$+$26 we again find a mean DM that is significantly smaller than
the value in the old catalog (based on the work of \cite{PnW-92}).  Our result 
is also discrepant from that of \cite{kuz_etal_98}, which is in good agreement 
with the old catalog value.  However, the new pulsar catalog 
\citep[][]{Hobbs_etal} cites value of 
$\rm 19.454 \pm 0.004 ~pc/cm^3$, which is fully consistent with our result.

For pulsar B0834$+$06 we find a $\langle DM \rangle$ ($\rm 12.867 \pm 0.002 ~pc/cm^3$) that is 
somewhat larger than the old catalog value of $\rm 12.8579 \pm 0.0002 ~pc/cm^3$ (based on the 
work of \cite{PnW-92}).  For this pulsar, \cite{kuz_etal_98} report a value 
of $\rm 12.865 ~pc/cm^3$, which agrees quite well with our result, whereas the new 
pulsar catalog \citep[][]{Hobbs_etal} cites a value of $\rm 12.889 ~pc/cm^3$, 
significantly higher than all the other numbers for this pulsar.

For B2016$+$28, our $\langle DM \rangle$ values (from 2 different pairs of frequencies) are 
consistent with each other, but are significantly smaller than the results cited
in the old catalog (based on the work of \cite{Craft-70}), the new catalog (based 
on the work of \cite{Hobbs_etal}), as
well as in \cite{kuz_etal_98}, all of which are consistent with each other.

Amongst all our results, the mean value of DM for PSR B1642$-$03 shows the largest 
discrepancy with the original catalog value of $\rm 35.665 \pm 0.005 ~pc/cm^3$ (based on very 
early work of \cite{Hunt-71}).  This is true for our DM results from both sets
of frequency pairs, though the discrepancy is more for our results obtained from
measurements at $\rm 325+610$ MHz bands (the difference in DM values from the two
frequency pairs is discussed separately in the next subsection).  We note that the 
DM of $\rm 35.73 ~pc/cm^3$ obtained by \cite{kuz_etal_98} is equally discrepant 
from this catalog value, and lies in between our two estimates.  A similar value 
($\rm 35.737 \pm 0.003 ~pc/cm^3$) is obtained from a multi-frequency timing analysis of over 
30 years of data for this pulsar by \cite{shaba_etal}.  The new pulsar catalog 
gives a value of $\rm 35.727 \pm 0.003 ~pc/cm^3$ \citep[][]{Hobbs_etal}, very close 
to the lower of our two results.
Clearly, either the original value of the DM reported for this pulsar was erroneously
estimated, or there has been a significant evolution of the DM of this pulsar from 
the early years of its discovery.

PSR B1642$-$03 is a particularly interesting pulsar, in several other respects.  
There is a significant uncertainty in the distance estimate to this pulsar.  The 
dispersion measure derived distance is $\rm 2.9 ~\rm kpc$ (with an uncertainty of 
$50 \%$), whereas the neutral hydrogen measurements provide a distance constraint 
of 160 pc (lower limit) \citep[][]{Graham_etal}. The smaller distance to this pulsar is also 
supported by a model \citep[by][]{PntH-69} that ascribes much of the DM to 
the presence of the $H_{II}$ region $\rm \zeta ~Oph.$ along the line of sight.  
Furthermore, \cite{shaba_etal} find that this pulsar has a very small proper 
motion and estimate transverse velocities of 2 and 30 $\rm km/s$ for the two distance 
estimates.  In addition, \cite{shaba_etal} also claim evidence for free precession 
in this pulsar, based on their analysis of the timing data.

Some of the above properties have interesting connections with the DM results.  For
example, \cite{gupta_etal_94} show that the observed scintillation properties of
this pulsar are consistent with a line of sight that goes through the limb of an 
$H_{II}$ region.  In such a case, a long term systematic variation of the pulsar
DM would be expected if there was sufficient transverse relative motion between the 
pulsar and the $H_{II}$ region. However we see no evidence for such a variation in
our data. On the other and, the observed DM 
changes could be part of a cyclic DM variation on large time scales, such as 
corresponding to the precession period ($\sim$ few 1000 days).  For example, results 
from the multi-frequency timing data of \cite{shaba_etal} show timing residuals 
at two different frequencies ($\sim$ 0.1 GHz and 0.6$-$2.3 GHz) which have differences 
between them that vary as a function of phase in the precession cycle.  The maximum 
amplitude of this difference is $\sim$ 1 ms, implying that the 0.1 GHz pulses arrive 
$\sim$ 1 ms later than the pulses at higher frequencies, at these phases.  One possible 
explanation of these variations is a cyclic change in DM of $\rm \sim 2.5 \times 10^{-3}
pc/cm^3$.  This, however, is too small compared to the changes and variations seen 
between the different DM values reported above.  Thus, although there is a lot more 
information about this pulsar, the nature and reason for the observed DM variations 
does not come out clearly.

From the results in this subsection, it is clear that constancy of DM estimates (at 
the level of 1 part in 1000 or better) for a pulsar can not be taken for granted.
Whether these small changes are due to genuine temporal evolution of pulsar DMs or
due to differences in the estimation techniques, remains to be established.

\subsection{DM values from different pairs of frequency bands} 
\label{subsec:DM_pairs}


For two of our pulsars -- B1642-03 and B2016+28 -- we carried out the observations 
at two pairs of frequency bands.  These data are almost simultaneous in that the 
observations at each epoch were taken within an hour or so of each other, and 
hence can be compared with each other.  PSR B1642$-$03 was observed at the frequency 
pairs of $\rm 325+610 ~MHz$ and $\rm 243+325 ~MHz$ and we find a significant difference 
in the mean DM values from these two sets of data (Table \ref{tab:dm_res}).  
The value obtained from the higher frequency combination ($\rm 325+610 ~MHz$) is 
higher than that obtained from the lower frequency combination ($\rm 243+325 ~MHz$).
On the other hand, for B2016$+$28 the DMs obtained from the two frequency pairs 
($\rm 325+610 ~MHz$ and $\rm 243+320 ~MHz$) are the same within errors, as determined by 
$\sigma_{DM_{\left(total\right)}}$.  

Though it is generally thought that the DM value for a pulsar is independent of the 
frequency of measurement, there have been reports in literature about differences 
in pulsar DMs that have been estimated from different parts of the radio spectrum 
(e.g. \cite{Shitov_etal}; \cite{Hankins-91}).  
In most of these results, the evidence is for an excess delay in the arrival of the 
pulses at low frequencies, when attempting to align them with a DM value computed
from the higher frequencies.  However, our results for PSR B1642$-$03, albeit for 
a relatively narrow range of radio frequencies, show an opposite trend in that
the DM value is larger for the higher frequency pair ($\rm 325+610 ~MHz$).  

There are different possible explanations for frequency dependent DM variations.  
For example, an evolution in the shape of the profile with frequency can play a 
role in changing the inferred alignment between the profiles at two different 
frequencies.  This should play a more significant role for pulsars with complex, 
multi-component profiles, but should be relatively insignificant for pulsars with
simple profiles (such as pulsars B1642$-$03 and B2016$+$28 in our sample).
Another interesting possibility is an extra time delay 
between emission received at two frequencies due to different heights of emission 
of these frequencies in the pulsar magneto-sphere \citep[e.g.][]{Kardashev_etal}, 
an idea that has not received much attention in the past.  These aspects will 
be examined in greater detail in a separate, forth-coming paper.

\subsection{DM differences from average profiles and single pulses}
\label{subsec:SP_AP_DMs}


As described in section \ref{sec:dm_est}, the DM estimates were obtained from 
two independent methods: measurement of delays between the average profiles (the 
AP method; results reported in Table \ref{tab:dm_res}) and measurement of delays 
between single pulse trains (the SP method).  We find, in general, that the DM 
results for a pulsar depend on the method of analysis.  For some pulsars, this 
difference is negligible, e.g. PSR B1642$-$03 (see Figure \ref{fig:dm_tm_srs}).  
For others it is significant: PSR B0329$+$54 is one such example in our study 
(see lower panel of Figure \ref{fig:dm_tm_srs}) -- the $\langle DM \rangle$ value obtained from 
the SP analysis is {\bf $\rm 26.7751 \pm 0.0007 ~ pc/cm^3$}, which is significantly
lower than the 
catalog value, which in turn is lower than the $\langle DM \rangle$ value from the AP analysis.

It is worth noting that dispersion measure values estimated from alignment of average 
profiles and those from cross-correlation of single pulse emission features have been
reported to be different in the past also.  \cite{Hankins-91} found that DM values 
from average profile measurements are significantly larger than those obtained from 
cross-correlation of pulsar micro-structure, for PSR B0950+08 and PSR B1133+16.  
\cite{sb_etal-92} have also investigated results for PSR B1133+16 over a ten year 
period, obtained using different techniques, and found significant variations in 
the DM values.

Further, as described in section \ref{sec:dm_est}, these two methods actually measure 
slightly different quantities.  Thus, the difference between average profile and single 
pulse analysis results that we find is not so surprising.  
A detailed description of these results and an investigation into 
the possible causes and implications of the same will be taken up in another forthcoming 
paper.

\subsection{Slow fluctuations of pulsar DMs}
\label{subsec:DM_fluc}


As mentioned at the beginning of this section, there is evidence for substantial temporal 
fluctuations in DM values for most of the pulsars.  A large part of this 
is likely to be due to the ISM.  A detailed study of this aspect will be 
taken up separately in another forthcoming paper.  Here, we briefly comment on the 
variations observed, comparing them with earlier published results.

Variations in pulsar DM, by definition, can arise due to either spatial and 
temporal changes in the electron density along the line of sight, or change in the 
distance to the pulsar, or both.  Electron density changes along the line of sight 
to the pulsar can be in the form of fluctuations resulting in DM fluctuations; 
alternatively, there can be a monotonic increase (or decrease) in DM due to the 
pulsar sampling a gradient of the electron density.  Most of our observed DM 
fluctuations (except PSR B2217$+$47) show fluctuations over a constant mean DM, 
indicating that the observed changes are due to electron density fluctuations in 
the ISM.  In the case of temporal change of distance to the pulsar with respect 
to the observer, the effect would manifest only as monotonic increase or decrease 
in pulsar DM. In our sample, PSR B2217$+$47 shows a monotonic increase of its DM 
(Figure \ref{fig:p2217}).  However, the amplitude of this change 
($\simeq 0.02 ~\rm pc/cm^3/year$) is such that it would require a very large radial
velocity ($\rm \sim 10^6 ~km/s$) through a normal density ISM ($\rm \sim 0.02 ~/cm^3$), 
or a very high density ISM ($\rm \sim 200 ~/cm^3$) for normal pulsar velocities 
($\rm \sim 100 ~km/s$).  
It is likely that the cause for this change is
due to the pulsar sampling an electron density gradient in the ISM, 
rather than due to radial motion of the pulsar. 

Such an electron density gradient can be produced by the line of sight to the 
pulsar crossing through a blob of enhanced plasma density.  Taking the electron 
density enhancement of $\Delta n_e \rm ~pc/cm^3$ in a wedge of thickness L pc, 
the observed change in DM is $\Delta DM = \Delta n_e . L$ .  The pulsar's
transverse displacement X (of $\rm \sim 3 \times 10^{-4} ~pc$), 
samples this electron density gradient in one year.  Assuming the wedge to 
be part of a spherical blob of radius L pc (and X $\sim$ L), we can estimate 
the electron density gradient to be $\sim 2 \times 10^5 ~\rm /cm^3/pc ~~or~~ 1 ~/cm^3/AU$.
This value is a lower limit -- if the cloud is closer to the observer, the electron 
density gradient could be even higher. 
Evidence for such AU-size clouds of enhanced electron density in the ISM also 
comes from scintillation observations of pulsars.  For example, \cite{bhat_etal-99} 
find evidence for clouds with length scales of $\sim 10$ AU and electron density
contrast $\rm \sim a ~few ~electrons/cm^3$.  Our results are similar, though a bit
on the higher side.  

Long term, slow DM variations, on time scales of weeks to months, have been studied 
in the past by \cite{Backer_etal} (3 pulsars) and \cite{PnW-91} (6 pulsars).  Whereas 
\cite{Backer_etal} report total DM fluctuations $\rm \sim 0.02 ~pc/cm^3$
over 1 $-$ 2 year periods, \cite{PnW-91} report typical variations $\rm \sim 0.002 ~pc/cm^3$
(and smaller) over similar time intervals.  Our results show $\sigma_{DM_{\left(total\right)}}$ 
 $\rm \sim 0.001$ to $0.007~ \rm ~pc/cm^3$ for most cases, implying total fluctuations \rm $\sim 
0.005$ to $0.03 ~\rm ~pc/cm^3$.  These are typically larger than those reported by \cite{PnW-91}, 
but comparable to the results of \cite{Backer_etal}.  

\subsection{Summary}
\label{subsec:summary}


We have presented a new experiment for accurate measurement of pulsar DMs using the
GMRT in a simultaneous, multi-frequency sub-array mode.  We have shown that single
epoch DM estimates using this technique can achieve an accuracy of 1 part in $10^4$ 
or better.  With improved sensitivity performance of the GMRT and faster sampling
that is now available, this accuracy can be improved in future experiments and the
technique can be extended to a larger set of pulsars.
From the large number of epochs of DM measurements for each of the 12 pulsars in
our sample, we are able to obtain fairly accurate estimates for the mean DM for most 
of them.  A detailed comparison of DM values in the literature with our mean DM
values highlights the lack of consistency (at the level of $\sim$ 1 part in 1000) 
in the different DM estimates, the reason for which remains to be understood.  We
have also briefly highlighted some of the other results from our data -- such as 
DM estimates from different frequency combinations, differences in average 
profile and single pulse DM values, and slow fluctuations of pulsar DMs with time
(which are most likely to be due to ISM effects) --  these will be the subject of 
follow-up papers.

\vspace{5mm}
\noindent{\large\bf Acknowledgments :} 

We thank the staff of the GMRT for help with the observations.  The GMRT is run 
by the National Centre for Radio Astrophysics of the Tata Institute of Fundamental 
Research.  We thank V. Kulkarni and R. Nityananda for help and encouragement during 
the initial stages of this work.  YG would like to acknowledge the help of P. Gothoskar
in the devising of the original experiment.


\begin{thebibliography}{}
\bibitem[Backer et al. (1993)]{Backer_etal} Backer, D. C., Hama, S., Hook, S. V., and Foster, R. S., 1993, {\it ApJ} {\bf 404}, 636.
\bibitem[Bartel et al. (1981)]{Bartel_etal} Bartel, N., Kardashev, N. S., Kuzmin, A. D.,
Nikolaev, N. Ya., Popov, M. V., Sieber, W., Smirnova, T. V., Soglasnov, V. A.,  and 
Wielebinski, R., 1981, {\it A\& A}, {\bf 93} 85.
\bibitem[Bhat et. al. (1999)] {bhat_etal-99} Bhat, N.D.R., Gupta, Y., Rao, P, 1999, {\it ApJ} {\bf 514}, 249.
\bibitem[Craft (1970)]{Craft-70} Craft, H. D., 1970, {\it PhD thesis}, {\bf Cornell University}.
\bibitem[Graham et al. (1974)] {Graham_etal} Graham, D.A., Mebold, U., Hesse, K.H., Hills, D.L., Wielebinski, R., 1974, {\it A\&A} {\bf 37}, 405.
\bibitem[Gupta et al. (1994)] {gupta_etal_94} Gupta, Y., Rickett, B. J., Lyne, A. G.,
1994, {\it MNRAS} {\bf 269}, 1035.
\bibitem[Gupta et al. (2000)]{gupta_etal_00} Gupta, Y., Gothoskar, P. B., Joshi, B. C., 
Vivekanand, M., Swain, R., Sirothia, S., and Bhat, N.D.R., 2000, in IAU Colloq. 177, 
Pulsar Astronomy, ed. M. Kramer, N. Wex, And R. Wielebinski (ASP Conf. Ser. 202; 
San Francisco: ASP), 277.
\bibitem[Hankins (1987)]{Hankins-87} Hankins, T.H. , 1987, {\it ApJ} {\bf 312}, 276.
\bibitem[Hankins (1991)]{Hankins-91} Hankins, T. H., Izvekova, V. A., Malofeev, V. M., 
Rankin, J. M., Shitov, Y. P., and Stinebring, D. R., 1991, {\it ApJ} {\bf 373}, L17. 
\bibitem[Hobbs et al. (2004)]{Hobbs_etal} Hobbs, G., Lyne, A. G., Kramer, M., 
Martin, C. E. and Jordan, C. A., 2004, {\it MNRAS} {\bf 353}, 1311.
\bibitem[Hunt (1971)]{Hunt-71} Hunt, G. C., 1971, {\it MNRAS}  {\bf 153}, 119.
\bibitem[Kardashev et al. (1982)]{Kardashev_etal} Kardashev, N. S., Nikolaev, N. Ya., 
Novikov, A. Yu., Popov, M. V., Soglasnov, V. A., Kuzmin, A. D., Smirnova, T. V., 
Bartel, N., Sieber, W., and Wielebinski, R., 1982, {\it A\& A}, {\bf 109} 340.
\bibitem[Kuzmin et al. (1998)]{kuz_etal_98} Kuzmin, A. D., Izvekova, V. A., Shitov, Yu. P., 
Sieber, W., Jessner, A., Wielebinski, R., Lyne, A. G., Smith, F. G., 1998, 
{\it A\&A Suppl. Ser.} {\bf 127}, 355.
\bibitem[Manchester \& Taylor (1972)]{MnT-72} Manchester, R. N., and Taylor, J. H., 
1972 {\it Ap Lett.} {\bf 10}, 67.
\bibitem[Phillips \& Wolszczan (1991)]{PnW-91} Phillips, J. A. and Wolszczan, A., 
1991, {\it ApJ} {\bf 382}, L27.
\bibitem[Phillips \& Wolszczan (1992)]{PnW-92} Phillips, J. A. and Wolszczan, A., 
1992, {\it ApJ} {\bf 385}, 273.
\bibitem[Prentice and ter Haar (1969)]{PntH-69} Prentice, A.J.R., Haar, D. ter, 1969, {\it MNRAS}
{\bf 146}, 423.
\bibitem[Shabanova et al. (2001)]{shaba_etal} Shabanova, T.V., Lyne, A.G.,
Urama, J.O., 2001, {\it ApJ} {\bf 552}, 321.
\bibitem[Shitov et al. (1988)]{Shitov_etal} Shitov, Yu. P., Malofeev, V. M. and 
Izvekova, V. A., 1988, {\it Sov. Astron. Lett.} {\bf 14(3)}, 181.
\bibitem[Stinebring et al. (1992)]{sb_etal-92} Stinebring, D.R., Thorsett, S. E., 
and Kaspi, V. M., 1992, in IAU Colloq. 128, ed. Hankins, T.H., Rankin, J.M., Gil, 
J.A., (Poland: Padagogical University Press), 349. 
\bibitem[Swarup et al. (1997)]{Swarup_etal} Swarup, G., Ananthakrishnan, S., Subrahmanya, 
C. R., Rao, A. P., Kulkarni, V. K., and Kapahi, V. K., 1997, in High Sensitivity Radio
Astronomy, ed. N. Jackson and R. J. Davis (Cambridge: Cambridge University Press).
\bibitem[Taylor, Manchester \& Lyne (1993)]{cat-93} Taylor, J. H., Manchester, R. N., 
and Lyne, A. G., 1993, {\it ApJS} {\bf 88}, 529.
\end{thebibliography}
\end{document}